\RequirePackage[2020-02-02]{latexrelease}
\documentclass[amssymb,prb,twocolumn,showpacs]{revtex4}
\usepackage{epsfig}
\usepackage{dcolumn}
\usepackage{amsmath}
\begin{document}
\title{\bf\ Quantum superposition in ultra-high mobility 2D photo-transport.}
\author{Jes\'us I\~narrea$^{1}$
}

\address{$^1$Escuela Polit\'ecnica
Superior,Universidad Carlos III,Leganes, Madrid, 28911, Spain.
}

\begin{abstract}
We investigate the striking properties that  magnetoresistance of irradiated two-dimensional electron systems presents
when their mobility is ultra-high ($\mu \gg 10^{7} cm^{2}V^{-1} s^{-1}$) and temperature is low ($T \sim 0.5$ K).
Such as, an abrupt
magnetoresistance collapse at low magnetic field and a resonance peak shift to the second harmonic ($2w_{c}=w$),
$w_{c}$ and $w$ being the cyclotron and radiation frequencies respectively.
We appeal to the principle of quantum superposition of
coherent states and  obtain that Schrodinger cat states (even and odd)
are key to explain magnetoresistance at these extreme mobilities.
On the one hand,  the Sch\"odinger cat states
system oscillates with $2w_{c}$, thus being
responsible of the resonance peak shift.
On the other hand, we obtain that
Schrödinger cat states-based scattering processes
give rise to a destructive effect when the odd states are involved, leading to a
 magnetoresistance collapse. The Aharonov-Bohm effect plays a central role in the latter, turning
 even cat states into odd ones.
We show that ultra-high mobility two-dimensional electron systems could
make a promising bosonic mode-based platform for quantum computing.


\end{abstract}
\maketitle
Two-dimensional electrons systems (2DES) under low
magnetic field ($B$), turn into a structure of coherent states\cite{prres} of the quantum harmonic oscillator.
The seminal idea of coherent states\cite{dodonov,yurke,noel,dodonov2,sro,glau,cohen} was introduced by Schrödinger\cite{sro} describing
minimum uncertainty constant-shape Gaussian wave packets of the quantum harmonic oscillator.
When irradiated under $B$ those systems give rise to the well-known
microwave-induced resistance oscillations (MIRO)\cite{mani,zudov} and zero resistance states (ZRS)\cite{mani,zudov}.
We studied these effects based on  {\it the microwave-driven electron orbit model}\cite{ina1,ina2,kerner,park,ina3,ina4}.
Thus, we found  that the time
it takes a scattered electron to jump between  coherent states, or evolution time $\tau$, equals
the cyclotron period, $T_{c}=2\pi/w_{c}$\cite{prres}. The rest of scattering processes
at different $\tau$ do not significantly contribute to the current.
In this way, MIRO reveal the presence of coherent states through this value
 for $\tau$ that is hidden in the peculiar positions that MIRO extrema
take in experiments\cite{mani,zudov}.
For instance, MIRO minima comply with $w/w_{c}=(j+1/4)$, $j$ being a positive integer.
 This is a universal result irrespective of platform and carrier.

Experimental evidences\cite{zudov2,rui,volkov} obtained with irradiated ultraclean ($\mu \gg 10^{7}$ $cm^{2}/Vs$) samples
of 2D electrons at low $T$ ($T\simeq 0.5 $ K),
show unexpected and striking MIRO features.
First, there is an almost complete magnetoresistance, ($R_{xx}$), collapse at low $B$,
observed also in the dark and well-known as {\it giant negative magnetoresistance}\cite{bock,kriisa,inagnm}.
Second, at high enough radiation power, there is a resonance peak shift from the expected position, $w_{c}=w$,  to
the second harmonic  $2w_{c}=w$, as if the system were under double $B$.
The latter is even more intriguing considering that MIRO, although with
lower intensity, keep showing up at the usual positions in smaller mobility samples.
We explain the above experiments based on
one of the most fundamental principles of quantum mechanics: the quantum superposition.
In our case the superposition of  coherent states giving rise to Schrodinger cat states.
According to the latter, ultra-clean 2DES  under a constant $B$ can be described by
a system  of Schrödinger cat states (even and odd) that oscillate with a frequency double ($2 w_{c}$) than expected
 considering the applied field ($ w_{c}$).
Thus, on the one hand, the double frequency oscillations explain the magnetoresistance resonance peak shift to $2w_{c}=w$.
On the other hand, we find that the $odd$ Schrödinger cat states when involved in scattering processes
undergo a destructive process that makes the scattering rate  and in turn $R_{xx}$, vanish.
The Aharonov-Bohm effect plays an essential role in the latter, transforming even Schrödinger cat states into odd ones when
a phase shift of $\pi$ is added.
Nevertheless, there is still a remanent part of $even$ Schrödinger cat states that would
be responsible of the experimentally obtained low intensity MIRO.
Thus, we conclude that ultra-high mobility 2DES under low $B$ and $T$, are made up of
even and odd Schrodinger cat states and then can become
a promising bosonic mode-based platform for quantum computing\cite{matos,cirac}.
Paradigmatic examples of quantum superposition are the even and odd
coherent states. These states are
superpositions of two coherent states of equal amplitude but separated
in phase by $\pi$ radians:\\
$|\alpha \rangle _{\binom{even}{odd}}=\frac{1}{2} N_{\binom{e}{o}}
\left[|\alpha \rangle \pm |-\alpha \rangle\right]$
where $N_{e}=\frac{e^{|\alpha|^{2}/2}}{\sqrt{\cosh(|\alpha|^{2})}}$ for even coherent states and
 $N_{o}=\frac{e^{|\alpha|^{2}/2}}{\sqrt{\sinh(|\alpha|^{2})}}$ for odd coherent states.
The plus sign corresponds to the even states and the minus to the odd ones.
The even and odd coherent states can be obtained from the quantum harmonic oscillator ground state with
the action of the even and odd displacement operators\cite{dodonov}, $D(\alpha_{\binom{even}{odd}})$.
The wave function for even and odd coherent states then reads,
 $\psi_{\alpha \binom{even}{odd}}=\langle x|D(\alpha_{\binom{even}{odd}})|\phi_{0}\rangle
=\frac{1}{2} N_{\binom{e}{o}}e^{i\vartheta_{\alpha}}\times$\\
$\left[e^{\frac{i}{\hbar}\langle p \rangle x}
\phi_{0}[x-X(0)-\langle x \rangle(t)] \pm e^{-\frac{i}{\hbar}\langle p \rangle x}
\phi_{0}[x-X(0)+\langle x \rangle(t)]\right]$
where $\phi_{0}$
is the ground state wave function of  the quantum harmonic oscillator, $\langle x \rangle(t)$ and $\langle p  \rangle(t)$ are the position and
momentum mean values respectively\cite{cohen},
$\langle x \rangle(t)=\sqrt{\frac{2\hbar}{m^{\ast} w_{c}}}|\alpha_{0}|\cos(w_{c}t)$
 and
$\langle p  \rangle(t)=-\sqrt{2m^{\ast}\hbar w_{c}}|\alpha_{0}|\sin(w_{c}t)$
where we  have used that  $\alpha=|\alpha_{0}| e^{-(iw_{c}t)}$.
\begin{figure}
\centering \epsfxsize=3.5in \epsfysize=2.in
\epsffile{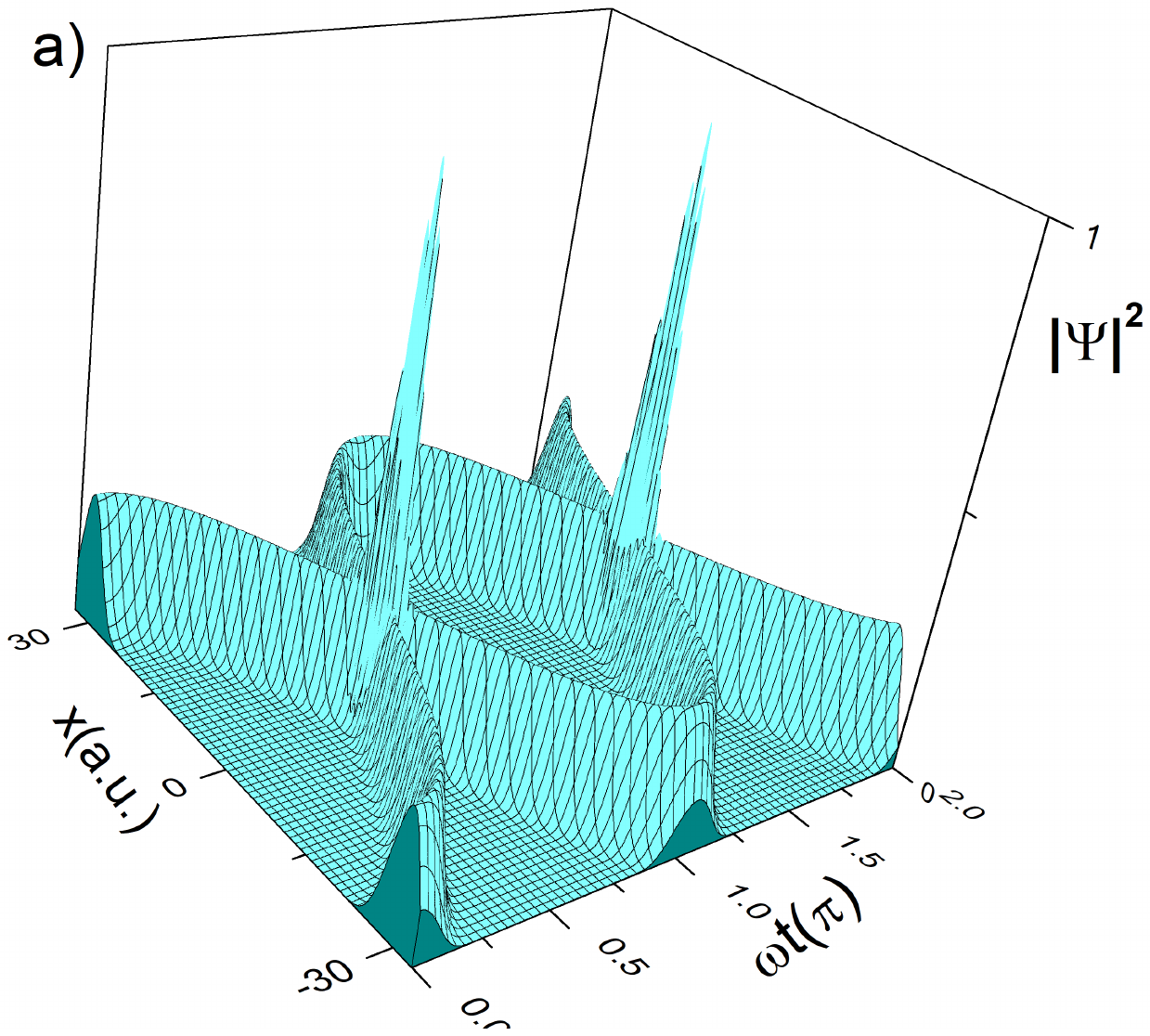}
\centering \epsfxsize=3.2in \epsfysize=2.in
\epsffile{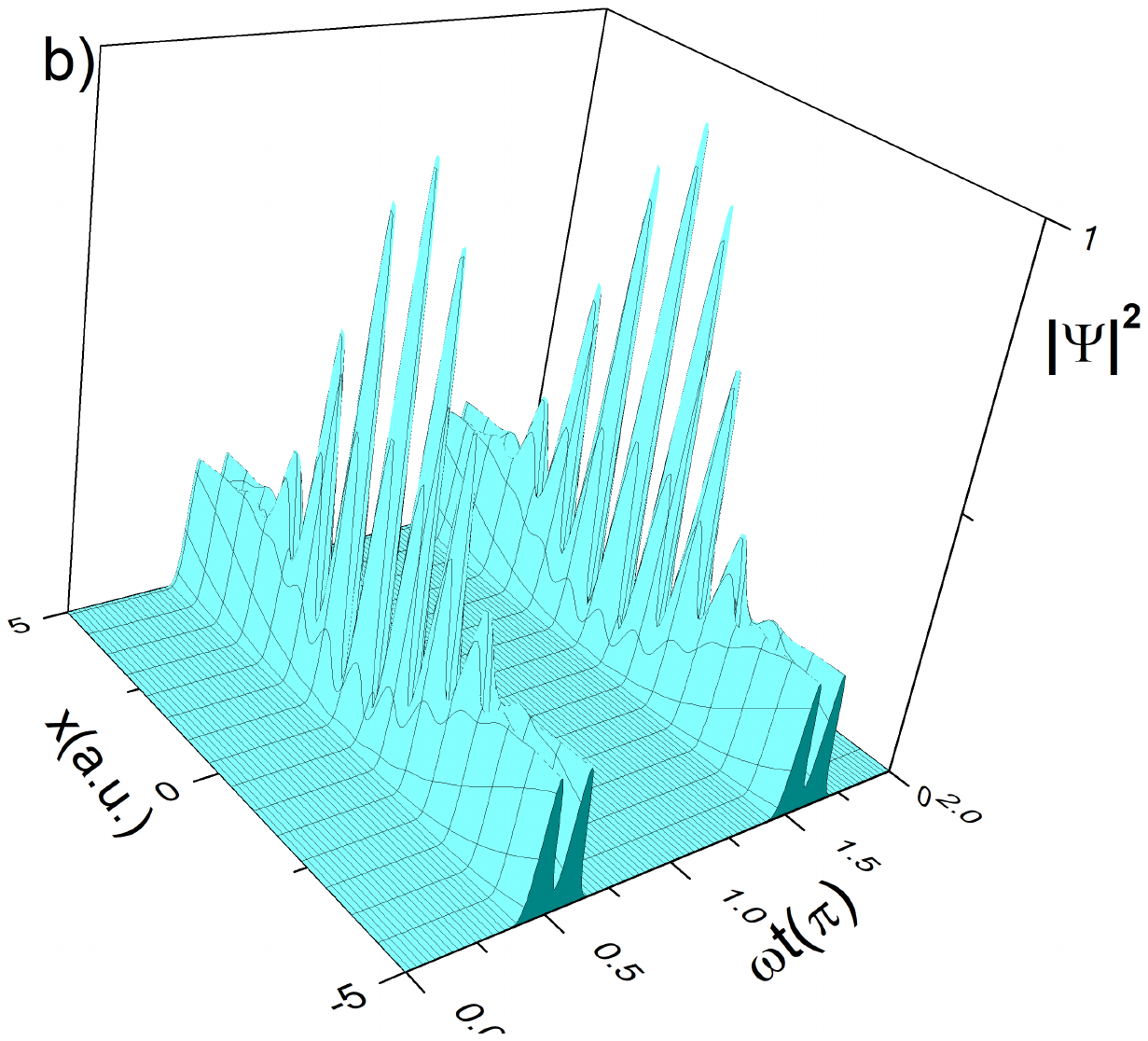}
\caption{a)  Even coherent state probability density. Quantum interference gives rise to
peaks at around  $w_{c}t=\pi/2$ and $3\pi/2$
 and $x  \sim 0 $. b) Zoom-in of the probability density around the peaks.
The probability densities shown are based on experimental values\cite{zudov2,rui}.
}
\end{figure}
When $|\alpha_{0}|$ is large, as in our case for low $B$, the coherent states $|\alpha \rangle$
and $ |-\alpha \rangle$ can be considered as macroscopically distinguishable.
Thus, the above superpositions are referred as {\it Schrodinger cat states}\cite{yurke,noel} and then
the normalization constants are
$N_{e}\simeq N_{o} \simeq 1/\sqrt{2}$.

Now we can calculate the probability density of the wave function $\psi_{\alpha \binom{even}{odd}}$,
\begin{eqnarray}
|\psi_{\alpha \binom{even}{odd}}|^{2}&=&\frac{1}{2}[
|\phi_{0}[x-X(0)-\langle x \rangle(t)|^{2}  \nonumber\\
&&+|\phi_{0}[x-X(0)+\langle x \rangle(t)|^{2}  \nonumber\\
&& \pm2\cos\left(\frac{2 x \langle p \rangle}{\hbar}\right)\left(\frac{m \hbar}{\pi w_{c}}\right)^{1/4} e^{-\frac{x^{2}+\langle x \rangle^{2}}{2(\Delta x)^{2}}} ]
\end{eqnarray}
where the last term gives rise to quantum interference.
Thus,  both types of superpositions
are made up of two Gaussian wave packets oscillating back and forth periodically
 with a phase difference of $\pi$ (see Figs. 1 and 2). When they cross, quantum interference shows up and peaks rise.
Besides,  the system as a whole oscillates with double frequency, $2w_{c}$.
Nevertheless, the most important point is the one of interference. This makes
the probability density peaks when $w_{c}t= \pi/2$ and  $3\pi/2$.
 Thus, we expect that the physical
processes of interest,  such as
electron scattering, will take place mainly at these points.
\begin{figure}
\centering \epsfxsize=3.5in \epsfysize=2.in
\epsffile{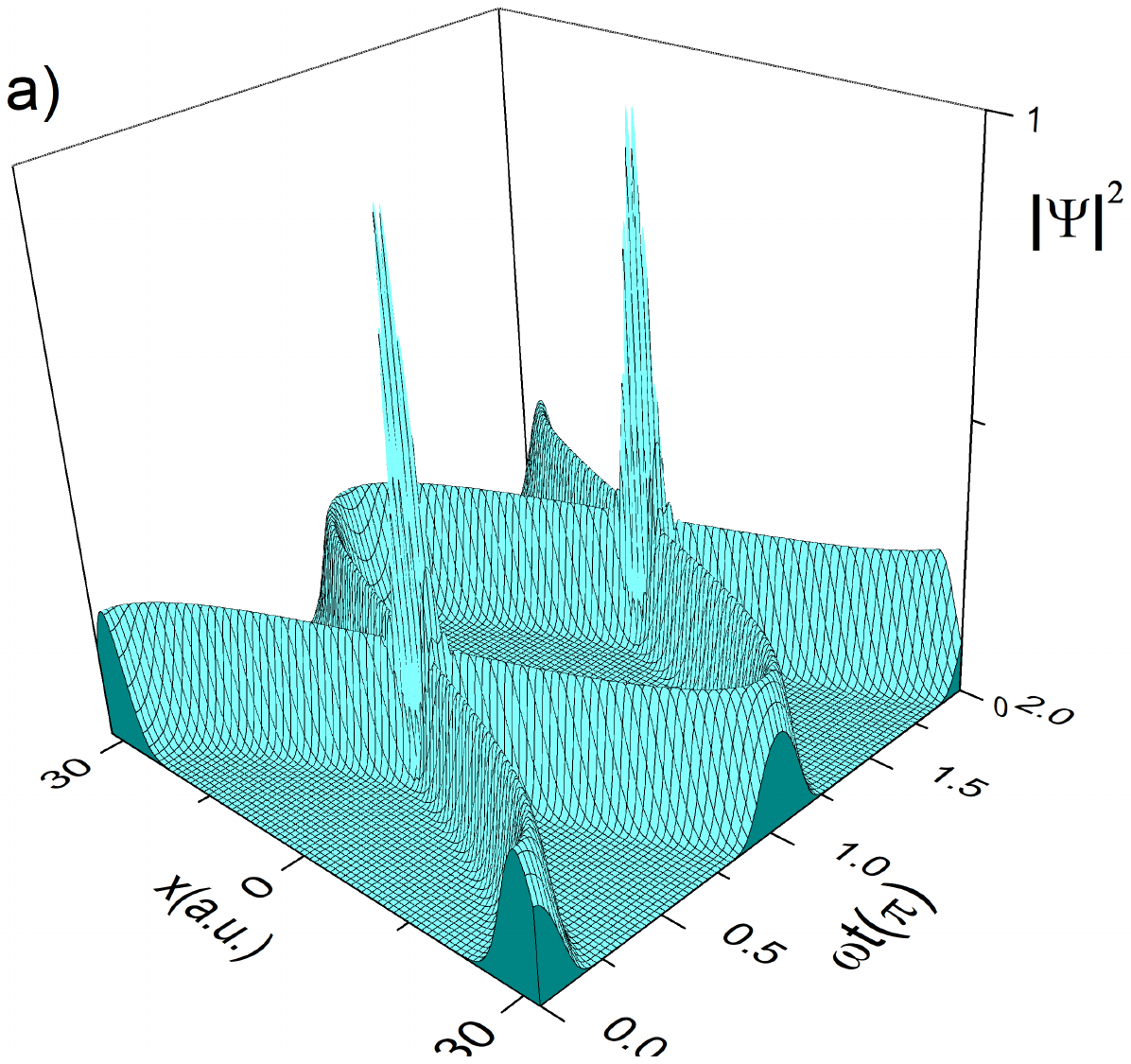}
\centering \epsfxsize=3.3in \epsfysize=2.in
\epsffile{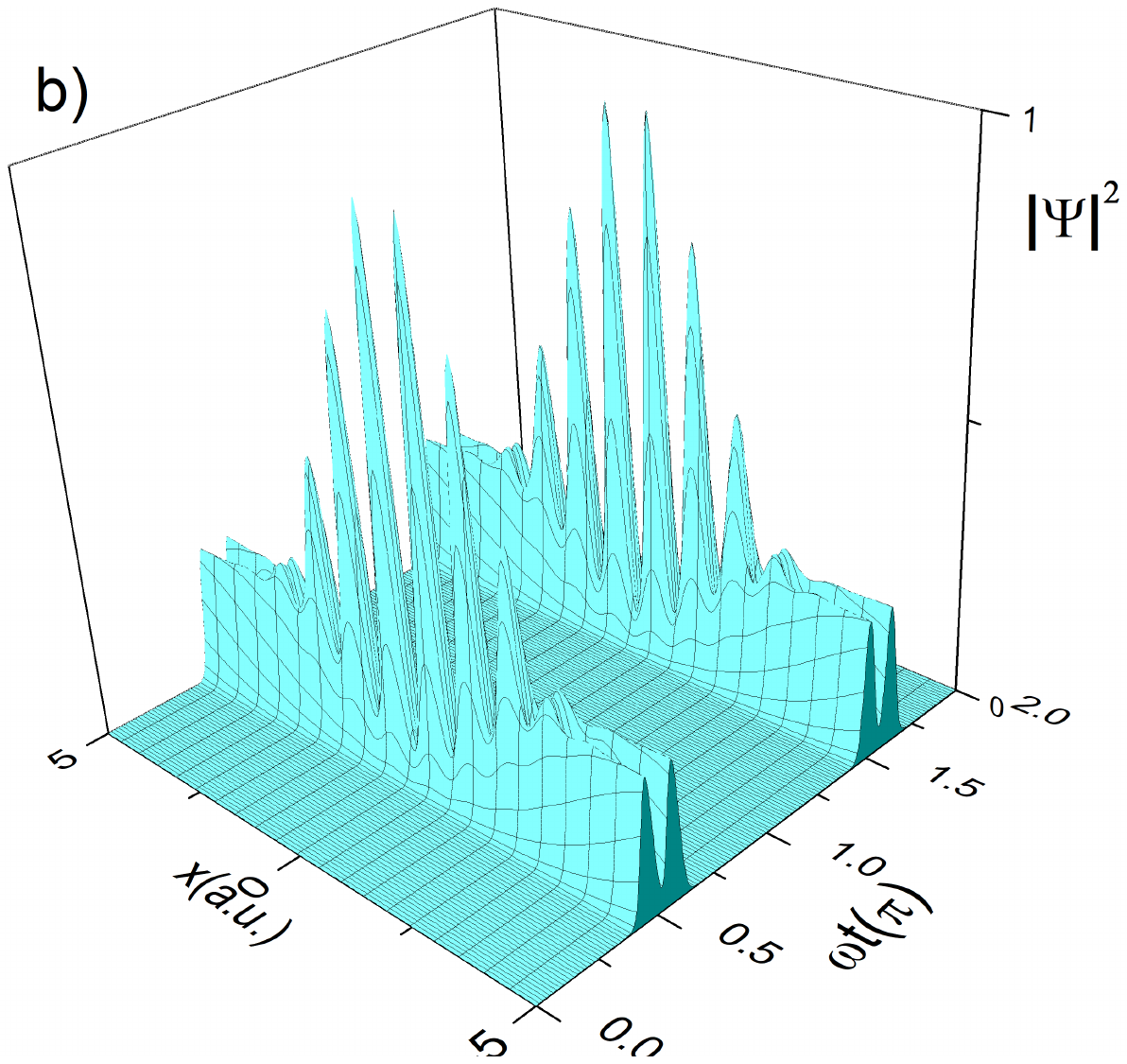}
\caption{Same as Fig. 1 for odd coherent states.
}
\end{figure}

To calculate  $R_{xx}$
 we apply a semiclassical Boltzmann model\cite{ridley,ando,askerov},
to obtain the longitudinal conductivity ${\sigma_{xx}}$, and then we calculate $R_{xx}$
 by the usual tensor relationships\cite{ina1,ridley,ando,askerov}.
Accordingly, ${\sigma_{xx}}$ is
proportional to the
the electron-charged impurities scattering rate\cite{ridley,ando,askerov} $W_{I}$ given by,
$  W_{I}=N_{i}\frac{2\pi}{\hbar}|<\psi_{\alpha^{'}}|V_{s}|\psi_{\alpha}>|^{2}\delta(E_{\alpha^{'}}-E_{\alpha})$
where $N_{i}$ is the number of charged impurities, $\psi_{\alpha}$ and $\psi_{\alpha^{'}}$ are the wave functions  corresponding to the initial and final cat states respectively,
 $V_{s}$ is the scattering potential for charged impurities\cite{ridley,ando,askerov}:
 $V_{s}=\sum_{q} V_{q}e^{i q_{x} x}$, and $q_{x}$ the $x$-component of $\overrightarrow{q}$,
the electron momentum change during the  scattering event.
The $V_{s}$ matrix element is given by\cite{ridley,ando,askerov}:
$|<\psi_{\alpha^{'}}|V_{s}|\psi_{\alpha}>|^{2}=\sum_{q}|V_{q}|^{2}|I_{\alpha,\alpha^{'}}|^{2}$
In the scattering rate  the essential part is
the integral $I_{\alpha,\alpha^{'}}$,  given by,
\begin{eqnarray}
&&I_{\alpha,\alpha^{'}}=\int^{\infty}_{-\infty} \psi_{ \alpha^{'}\binom{e}{o}} e^{i q_{x} x}\psi_{\alpha \binom{e}{o}} dx=\nonumber\\
\large
&&\frac{N^{2}_{\binom{e}{o}}}{2}\int^{\infty}_{-\infty}[e^{\frac{-ix(\langle p'\rangle-\langle p\rangle)}{\hbar}} e^{-\frac{(x-X^{'}_{0}-\langle x^{'}\rangle)^{2}}{2(\Delta x)^{2}}}e^{-\frac{(x-X_{0}-\langle x\rangle)^{2}}{2(\Delta x)^{2}}}\nonumber\\
&&+e^{\frac{ix(\langle p'\rangle-\langle p\rangle)}{\hbar}} e^{-\frac{(x+X^{'}_{0}+\langle x^{'}\rangle)^{2}}{2(\Delta x)^{2}}}e^{-\frac{(x-X_{0}+\langle x\rangle)^{2}}{2(\Delta x)^{2}}}\nonumber\\
&&\pm e^{\frac{-ix(\langle p'\rangle+\langle p\rangle)}{\hbar}} e^{-\frac{(x-X^{'}_{0}-\langle x^{'}\rangle)^{2}}{2(\Delta x)^{2}}}e^{-\frac{(x-X_{0}+\langle x\rangle)^{2}}{2(\Delta x)^{2}}}\nonumber\\
&&\pm e^{\frac{ix(\langle p'\rangle+\langle p\rangle)}{\hbar}} e^{-\frac{(x-X^{'}_{0}+\langle x^{'}\rangle)^{2}}{2(\Delta x)^{2}}}e^{-\frac{(x-X_{0}-\langle x\rangle)^{2}}{2(\Delta x)^{2}}}]dx
\end{eqnarray}
where $\langle p^{'}  \rangle(t^{'})=-\sqrt{2m^{\ast}\hbar w_{c}}|\alpha_{0}|\sin(w_{c}t')$, $\langle x' \rangle(t')=\sqrt{\frac{2\hbar}{m^{\ast} w_{c}}}|\alpha_{0}|\cos(w_{c}t')$ and $t'=t+\tau$. $t$ being the scattering initial time and $t'$ the final scattering time.
As we said above, according to the probability density the scattering processes will take place more likely
 at the probability peaks.
Thus, for instance, for the
case of $w_{c}t=\pi/2$ we get to,
\begin{eqnarray}
&&I_{\alpha,\alpha^{'}}=e^{i q_{x} \frac{X^{'}_{0}+X_{0}}{2}} e^{-\frac{q_{x}^{2}(\Delta x)^{2}}{2}} \times\nonumber\\
&& [e^{i q_{x}\frac{1}{2}\sqrt{\frac{2\hbar}{mw_{c}}}|\alpha_{0}|\cos(\frac{\pi}{2}+w_{c}\tau)}
e^{-\frac{[\Delta X_{0}+\sqrt{\frac{2\hbar}{mw_{c}}}|\alpha_{0}|\cos(\frac{\pi}{2}+w_{c}\tau)]^{2}}{8(\Delta x)^{2}}}\nonumber\\
&&+e^{-i q_{x}\frac{1}{2}\sqrt{\frac{2\hbar}{mw_{c}}}|\alpha_{0}|\cos(\frac{\pi}{2}+w_{c}\tau)}
e^{-\frac{[\Delta X_{0}-\sqrt{\frac{2\hbar}{mw_{c}}}|\alpha_{0}|\cos(\frac{\pi}{2}+w_{c}\tau)]^{2}}{8(\Delta x)^{2}}}\nonumber\\
&&\pm e^{i q_{x}\frac{1}{2}\sqrt{\frac{2\hbar}{mw_{c}}}|\alpha_{0}|\cos(\frac{\pi}{2}+w_{c}\tau)}
e^{-\frac{[\Delta X_{0}+\sqrt{\frac{2\hbar}{mw_{c}}}|\alpha_{0}|\cos(\frac{\pi}{2}+w_{c}\tau)]^{2}}{8(\Delta x)^{2}}}\nonumber\\
&&\pm e^{-i q_{x}\frac{1}{2}\sqrt{\frac{2\hbar}{mw_{c}}}|\alpha_{0}|\cos(\frac{\pi}{2}+w_{c}\tau)}
e^{-\frac{[\Delta X_{0}-\sqrt{\frac{2\hbar}{mw_{c}}}|\alpha_{0}|\cos(\frac{\pi}{2}+w_{c}\tau)]^{2}}{8(\Delta x)^{2}}}]
\nonumber\\
\end{eqnarray}
Remarkably enough, due to the large value of $|\alpha_{0}|$ the above expression turns out to
be negligible (real exponentials tend to zero\cite{prres}) except when $\tau=\frac{2\pi}{w_{c}}$ or $\tau=\frac{2\pi}{2 w_{c}}$.
Then, taking this into account,
we get to the final expression,
\begin{eqnarray}
I_{\alpha,\alpha^{'}}&=&2 e^{i q_{x} \frac{X^{'}_{0}+X_{0}}{2}} e^{-\frac{[\Delta X_{0}]^{2}}{8(\Delta x)^{2}}-\frac{q_{x}^{2}(\Delta x)^{2}}{2}}
 \nonumber\\
&& \pm 2 e^{i q_{x} \frac{X^{'}_{0}+X_{0}}{2}} e^{-\frac{[\Delta X_{0}]^{2}}{8(\Delta x)^{2}}-\frac{q_{x}^{2}(\Delta x)^{2}}{2}}
\end{eqnarray}
Then, for  the even Schr\"odinger cat states we obtain, $ I_{\alpha,\alpha^{'}}=4 e^{i q_{x} \frac{X^{'}_{0}+X_{0}}{2}} e^{-\frac{[\Delta X_{0}]^{2}}{8(\Delta x)^{2}}-\frac{q_{x}^{2}(\Delta x)^{2}}{2}}$
whereas for the odd ones $ I_{\alpha,\alpha^{'}}=0$.
 We also obtain, not shown, that when one of the
scattering-involved cat states, initial or final, is odd the
scattering integral is zero too, i.e.,
\begin{equation}
I_{\alpha,\alpha^{'}}=\int^{\infty}_{-\infty} \psi_{ \alpha^{'}\binom{e}{o}} e^{i q_{x} x}\psi_{\alpha\binom{o}{e}}dx=0
\end{equation}

At this point is where a geometrical phase shift similar to the Aharonov-Bohm effect comes into play.
Starting from an even cat state, $(|\alpha\rangle + |-\alpha\rangle )$,
we consider that both $|\alpha\rangle$  and  $|-\alpha\rangle $,  are under the same vector potential $\vec{A}$ that is taken
as constant. Thus, one of the sates, say $|\alpha\rangle$, in its motion sees $\vec{A}$ in opposite
direction on average (see Fig. 3). However the other state, $|-\alpha\rangle$, that is  $\pi$ radians delayed, displaces with the same direction as $\vec{A}$
giving rise to a phase shift between them.  This scenario is similar to the Aharonov-Bohm effect but now $B$ is present
in the two wave packet trajectories.
Therefore, the individual states $|\alpha\rangle$ and $|-\alpha\rangle$,
in
their time evolution, after half a cyclotron period $T_{c}$ jointly
complete a circle encompassing a magnetic flux $\Phi=B\times \pi R_{c}^{2}$ (see Fig. 3).
 Where $ R_{c}=l_{B}\sqrt{2(|\alpha|^{2}+1/2)}$  \cite{cohen},
$l_{B}$ being the magnetic length, $l_{B}=\hbar /eB$.
Thus, the relative geometrical phase shift acquired  is $\Delta\phi =\frac{e}{\hbar}\oint \vec{A}\cdot d\vec{r}= 2\pi \Phi/\Phi_{0}$ where $\Phi_{0}=e/h$ is the
flux quantum. Then the Schrödinger cat state becomes $|\alpha\rangle + e^{i\Delta\phi} |-\alpha\rangle $.
Considering that $|\alpha|^{2}=\langle n \rangle$, $n$ being a positive integer (Landau level index), a straightforward
calculation leads us to  $\Delta\phi = 2\pi \langle n \rangle + \pi$.
Thus, $\Delta\phi$ is independent of the magnetic field intensity an in turn of $\Phi$.
This result contrasts to the Aharonov-Bohm effect where the shift depends on  $\Phi$.
Accordingly, after $T_{c}/2$ an even state  turns into an odd one and viceversa.
Therefore, over a complete oscillation the Schrödinger cat state is half of the time even and half
of the time odd alternating the parity.
Then, if the cat states scatter when in the odd half of the oscillation,  the scattering suffers
a destructive interference effect. In the same way as when the state scatters in the even part
but ends up in an odd state.
 In a stationary scenario  such  processes give rise to
the $R_{xx}$ collapse as experiments show.

\begin{figure}
\centering \epsfxsize=3.5in \epsfysize=1.8in
\epsffile{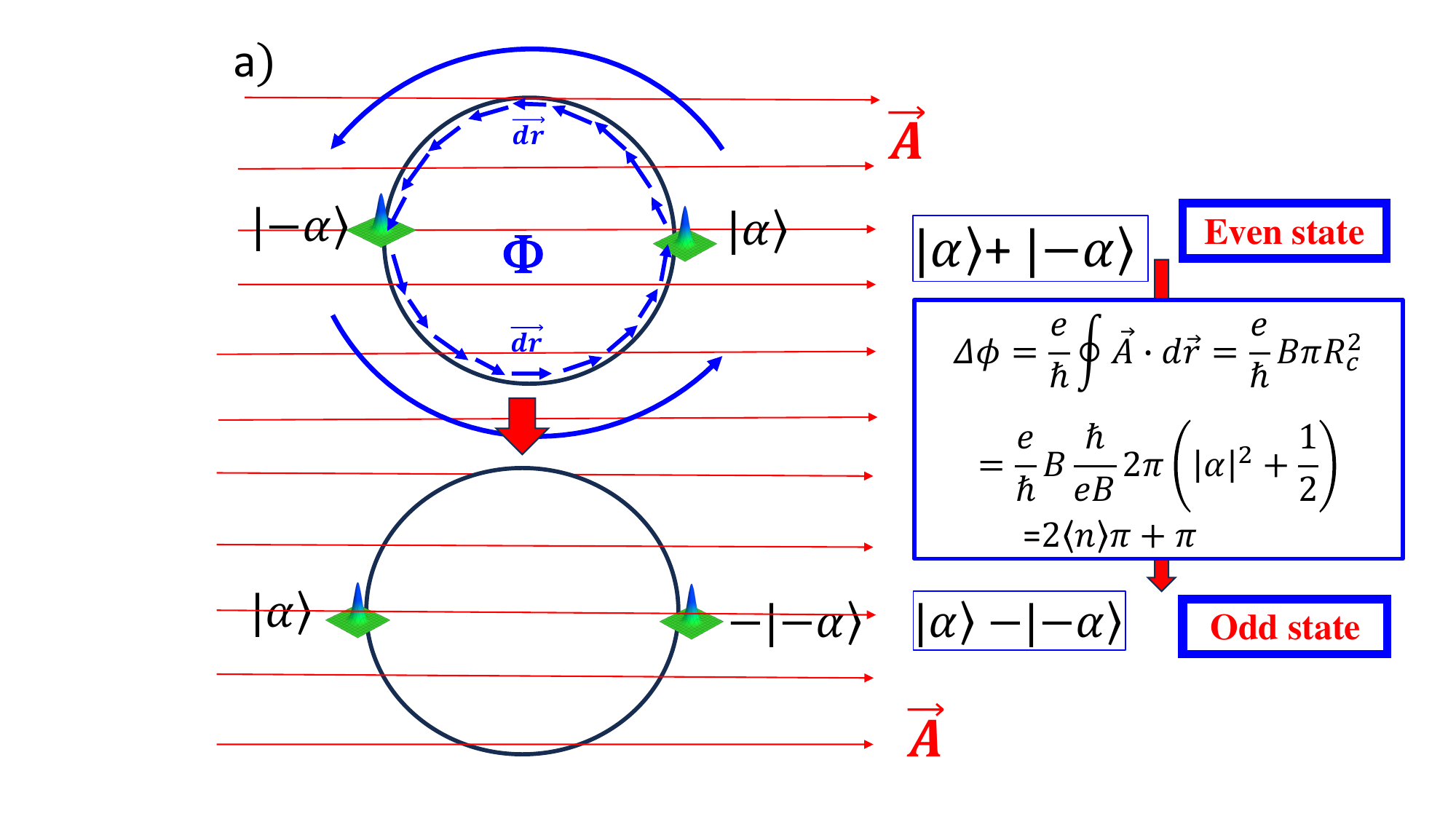}
\centering \epsfxsize=1.8in \epsfysize=0.5in
\epsffile{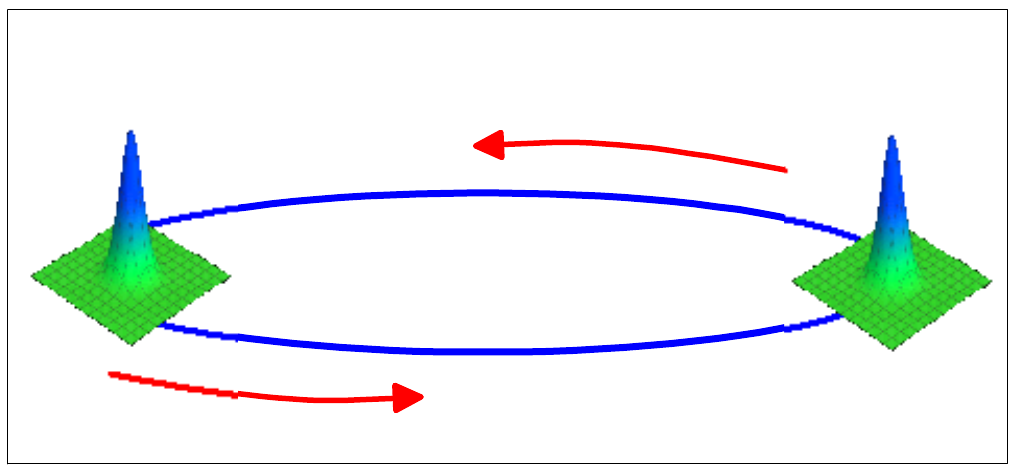}
\centering \epsfxsize=3.3in \epsfysize=1.8in
\epsffile{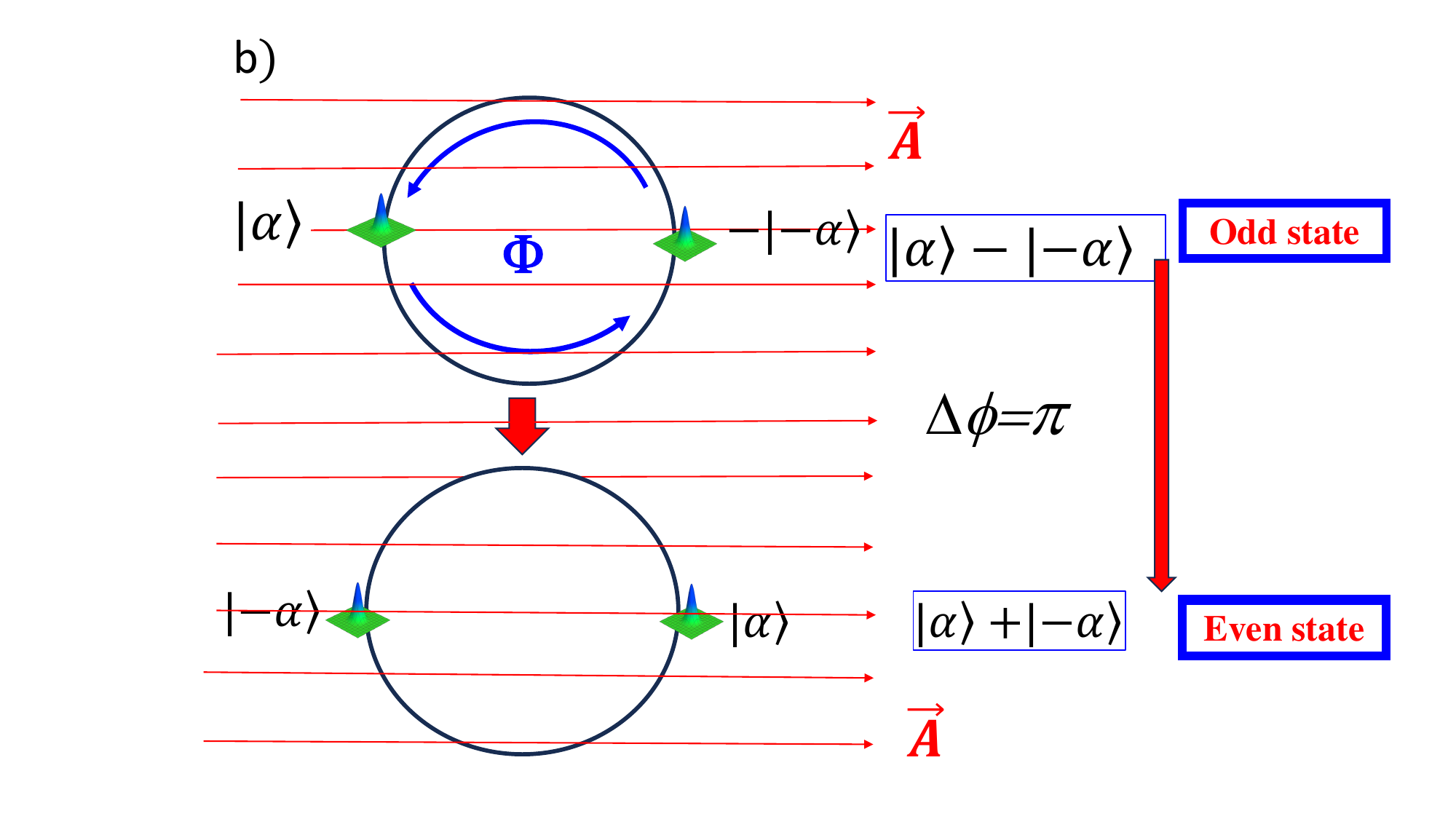}
\caption{a)  Schematic diagram of the even to odd Schrödinger cat state
 transition based on a phase shift. b) Same as in a) for the odd to even transition.
 Inset shows an even or odd Schrödinger cat state, with the two Gaussian wave packets with a delay of $\pi$ radians.
For both panels, the circles correspond to  the classical trajectories followed by
the two wave packets of the Schrödinger cat states. }
\end{figure}


 Thus, quantum superposition and quantum interference along with a phase shift (similar to the Aharonov-Bohm effect)
 are essential to explain the abrupt $R_{xx}$ collapse in ultra-high mobility 2DES.
But further than that, we can identify a novel bosonic mode-based platform to realize robust
even and odd Schr\"odinger cat states: the ultra-high mobility 2D
electron systems at low $B$.
Taking all the above into account, the Schrodinger cat states-based  magnetoresistance of 2DES
can be theoretically described by   individual $2w_{c}$-coherent states  with
two  $\tau$ values, $\frac{2\pi}{2w_{c}}$ and  $\frac{2\pi}{w_{c}}$.
This has to be reflected in their scattering-dependent physical properties such as
 MIRO.
Then, applying  the microwave-driven electrons orbits model \cite{ina1,ina2} to this novel scenario\cite{prres} we obtain an expression for
the irradiated $R_{xx}$ given by,
\begin{equation}
R_{xx}\propto \frac{e E_{o}}{m^{*}\sqrt{(2w_{c})^{2}-w^{2})^{2}+w^{2}\gamma^{2}}}\left(\sin w \frac{2\pi}{w_{c}}+\sin w \frac{2\pi}{2w_{c}}\right)
\end{equation}
The sum between brackets in the  $R_{xx}$ expression is made up of two contributions. The first corresponds to
$\tau=\frac{2\pi}{w_{c}}$ and then the cat state parity is conserved during the scattering.
The second part corresponds to $\tau=\frac{2\pi}{2w_{c}}$, and accordingly, during scattering the parity changes from
even to odd or from odd to even, giving a null scattering integral. Thus the second contribution can be ruled out and
the $R_{xx}$ expression finally reads,
$R_{xx}\propto \frac{e E_{o}}{m^{*}\sqrt{(2w_{c})^{2}-w^{2})^{2}+w^{2}\gamma^{2}}}\left(\sin w \frac{2\pi}{w_{c}}\right)$.
This expressions explains first the resonance peak shift with the amplitude denominator and the MIRO extrema positions with
the sine term. Recall  the current is held only by even states under irradiation and  in the dark .
\begin{figure}
\centering \epsfxsize=3.in \epsfysize=1.5in
\epsffile{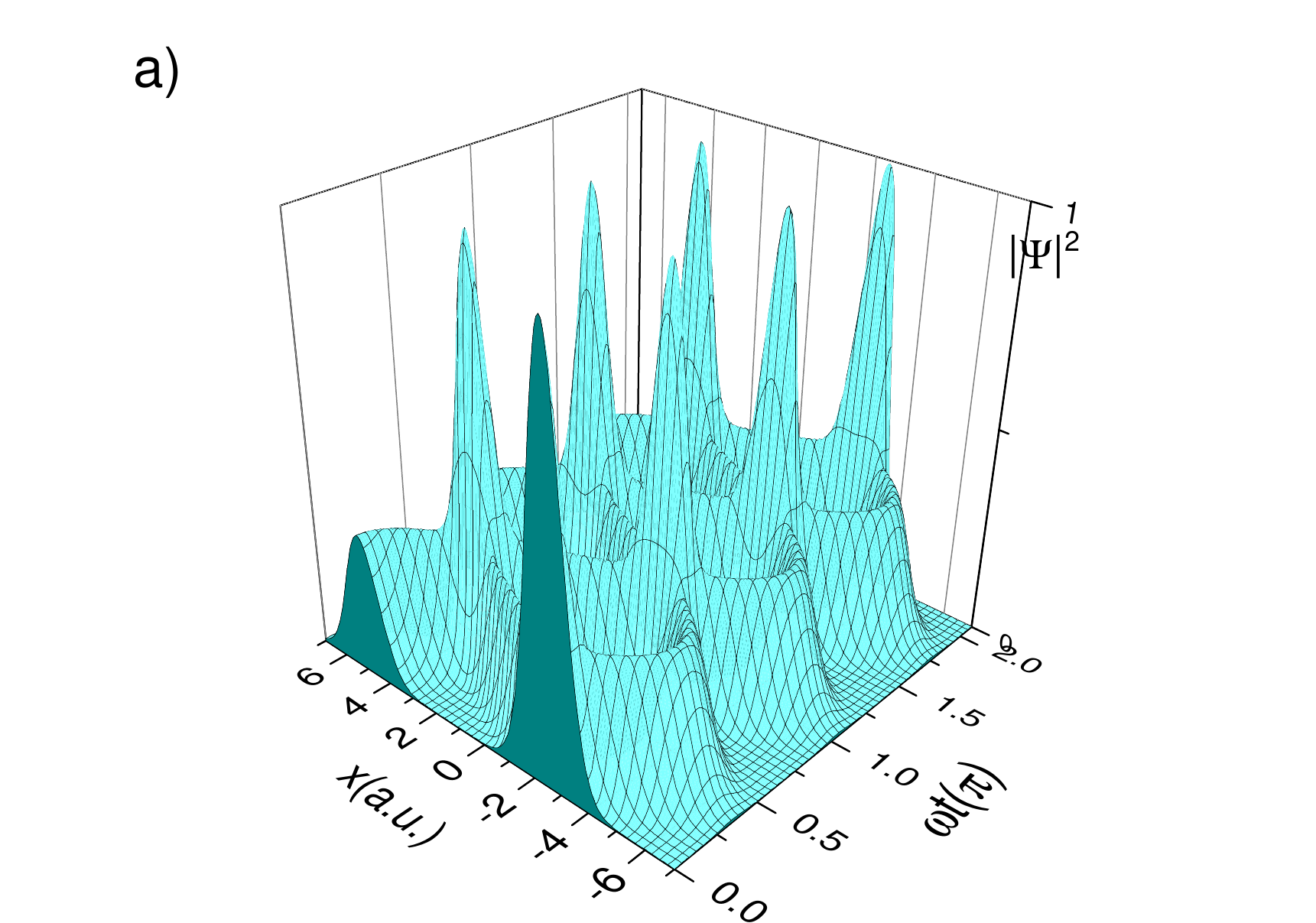}
\centering \epsfxsize=2.5in \epsfysize=1.2in
\epsffile{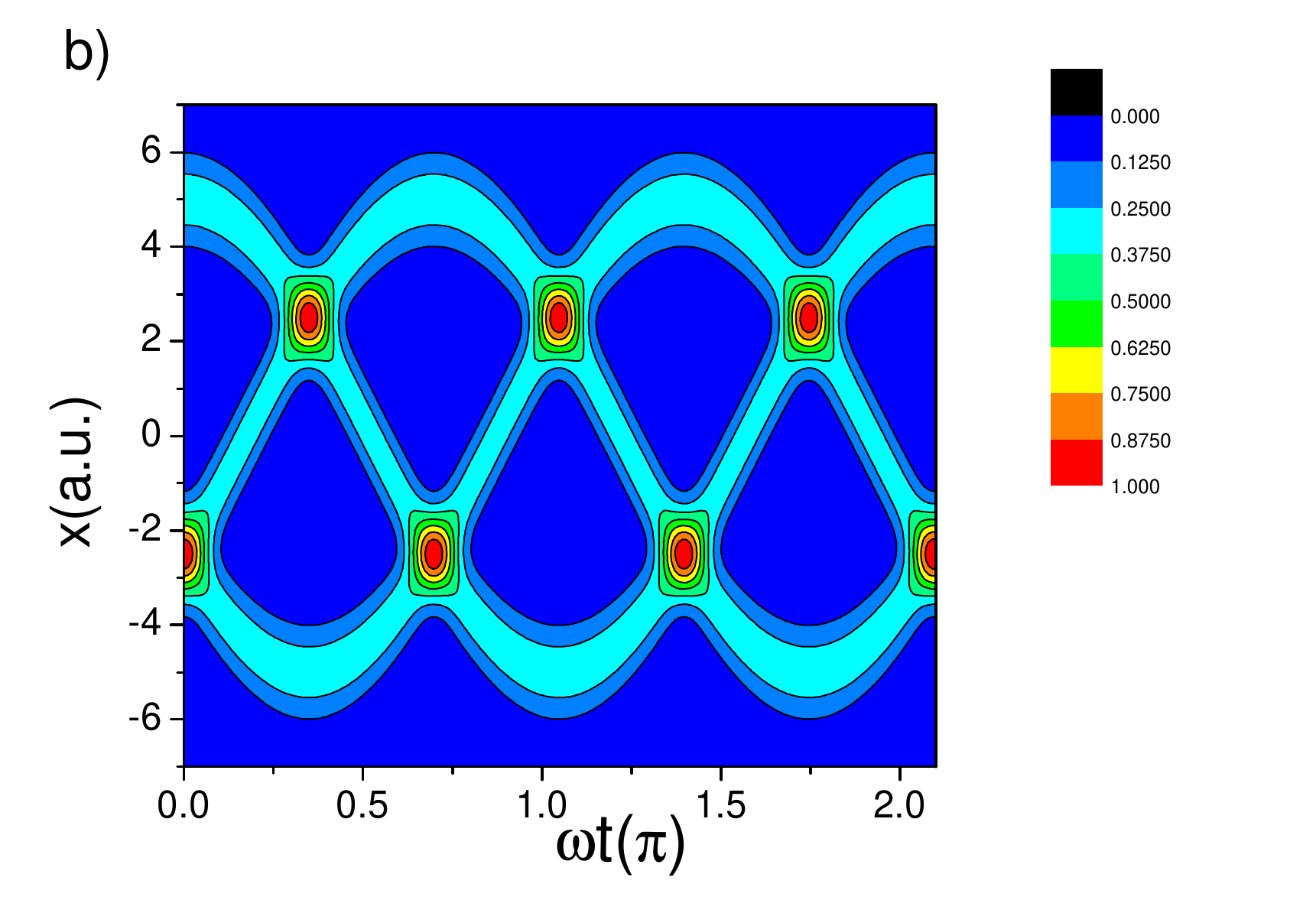}
\caption{a) Probability  density of a three component Schrodinger cat
state. b) Contour plot of the state in a).}
\end{figure}
\begin{figure}
\centering \epsfxsize=3.in \epsfysize=4.in
\epsffile{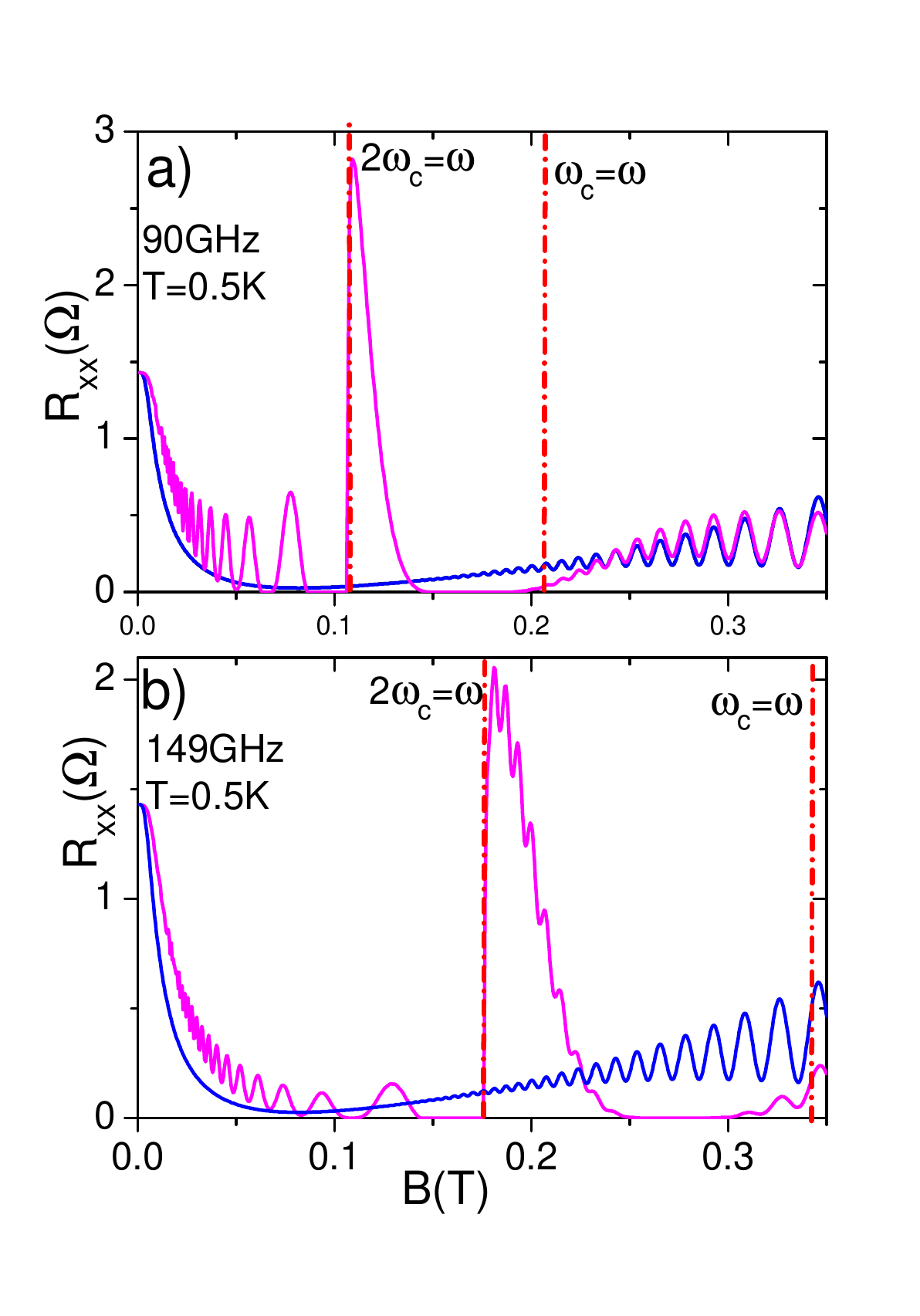}
\caption{a) Calculated magnetoresistance vs $B$  for two-component cat states. The radiation frequency is $90$ GHz and $T=0.5$ K (magenta curve).
We also exhibit dark magnetoresistance (blue curve). When radiation is on, the curve shows MIRO, zero resistance states
and the rise of a shifted resonance peak at $2w_{c}=w$.
b) Same as  in a) for a radiation frequency of  $149$ GHz}
\end{figure}
\begin{figure}
\centering \epsfxsize=3.3in \epsfysize=2.5in
\epsffile{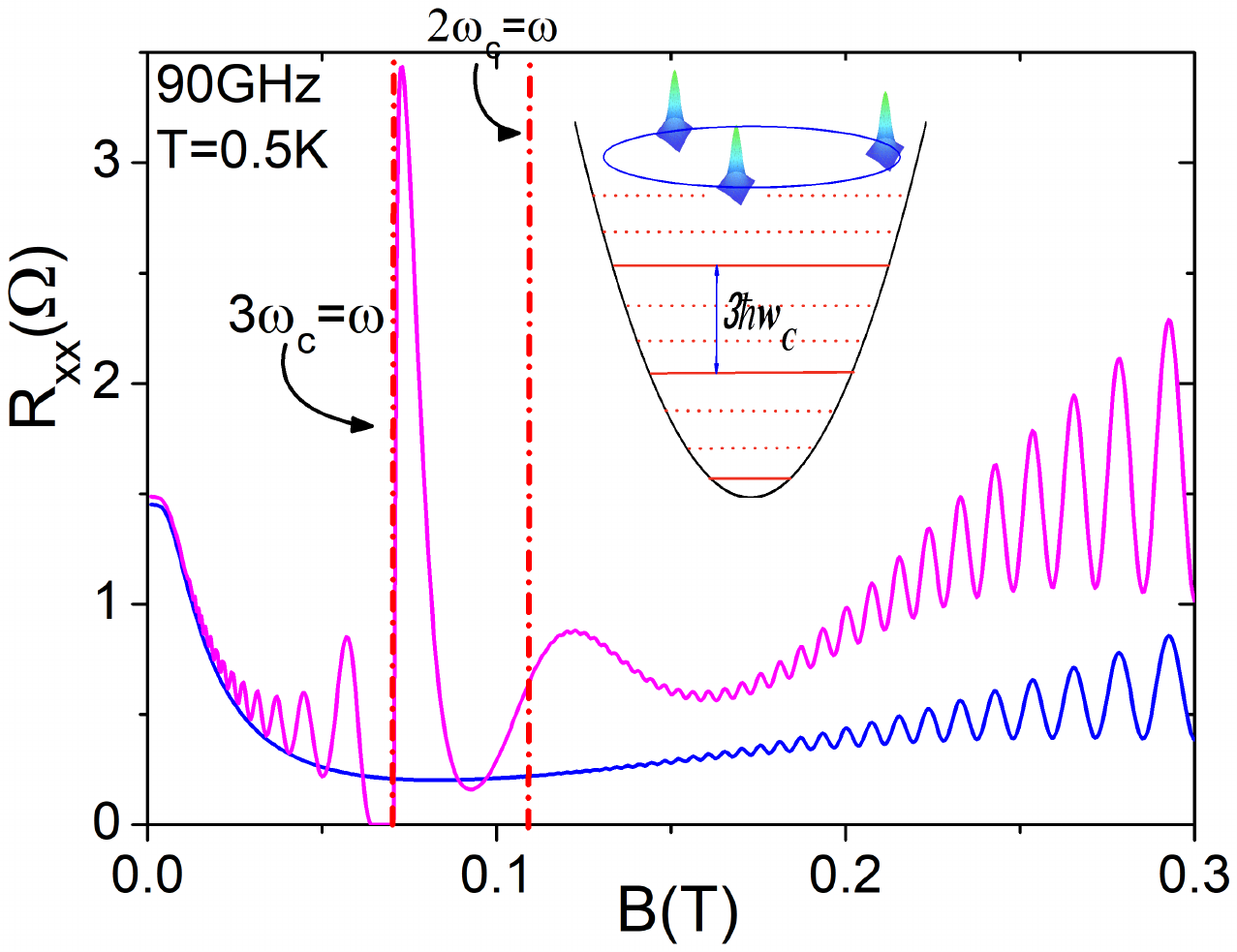}
\caption{Calculated magnetoresistance vs $B$  for three-component cat states.  The radiation frequency is $90$ GHz and $T=0.5$ K.
Dark magnetoresistance (blue curve) is also exhibited.
The irradiated curve (magenta curve) shows MIRO, zero resistance states
and the rise of a shifted resonance peak at $3w_{c}=w$.
Inset shows a schematic diagram of a three-component cat state.}
\end{figure}

We can extend the above theory to Schrodinger cat states with  three components\cite{vlastakis} with
a state expression that reads:
$|\alpha _{3} \rangle =\frac{1}{\sqrt{3}}  \left[|\alpha \rangle \pm |e^{i2\pi/3}\alpha \rangle \pm |e^{i4\pi/3}\alpha \rangle \right]
=\sqrt{3} e^{-|\alpha|^{2}/2}\sum_{n}\frac{\alpha^{3n}}{\sqrt{(3n)!}}|\phi_{3n}\rangle$
with  an energy of $E_{3n}=\hbar w_{c}(3n+1/2)$ and an energy difference between
levels of
$\Delta E_{3n}=3 \hbar w_{c}$. Therefore, one out of three levels is populated.
We can obtain the expression of the wave function similarly as the two component cat state
giving:
\begin{eqnarray}
\psi_{\alpha_{3}}(x,t)=\frac{1}{\sqrt{3}} e^{i\vartheta_{\alpha}}
[e^{\frac{i}{\hbar}\langle p \rangle(t) x}
\phi_{0}[x-X(0)-\langle x \rangle(t)] \nonumber\\
 \pm e^{-\frac{i}{\hbar}\langle p_{1} \rangle(t) x}
\phi_{0}[x-X(0)+\langle x_{1} \rangle(t)\nonumber\\
\pm e^{-\frac{i}{\hbar}\langle p_{2} \rangle(t) x}
\phi_{0}[x-X(0)+\langle x_{2} \rangle(t)]]
\end{eqnarray}
where $\langle x_{1} \rangle(t)$ and $\langle x_{2} \rangle(t)$ have similar expression to
$\langle x \rangle(t)$ with a phase difference of $2\pi/3$ and $4\pi/3$ respectively.
Similar conditions apply to $\langle p_{1}\rangle(t)$ and $\langle p_{2}\rangle(t)$ regarding $\langle p\rangle(t) $.
Accordingly we calculate the probability density obtaining interference terms
that peak now at $w_{c}t=0,\frac{\pi}{3},\frac{2 \pi}{3},\frac{3\pi}{3},\frac{4\pi}{3},\frac{5\pi}{3}$
and $2\pi$ (see Fig. 4). We observe that the system now oscillates with $3w_{c}$.
Similarly as the even and odd cat states, we calculate with a semiclassical Boltzman model
 dark
and irradiated $R_{xx}$. After similar algebra as before, we obtain that $R_{xx}$
is negligible except when $\tau$ is given by, $\tau= \frac{2\pi}{3w_{c}}$,  $\frac{4\pi}{3w_{c}}$ and $\frac{6\pi}{3w_{c}}$.
Remarkably enough, when irradiated we obtain a resonance peak shift at $3w_{c}=w$, but still keeping MIRO
their usual positions (see Fig. 6). This is what would be observed in MIRO experiments with even higher mobility  ($\mu \sim 10^{8} cm^{2}V^{-1} s^{-1}$) samples.

In Fig. 5, we exhibit calculated results for  dark and  irradiated
$R_{xx}$ vs $B$  of a two-component cat state. The  radiation frequency is (a) $90$ GHz and (b) $149$ GHz at  $T=0.5$ K.
The curves show a surprising strong $R_{xx}$ collapse or giant negative magnetoresistance.
When radiation is on, the curves also show MIRO and ZRS, but the most
striking effect is the rise of a shifted resonance peak at $2w_{c}=w$ instead of
the expected position, $w_{c}=w$.
However, MIRO extrema show up at the ususal positions in agreement with
MIRO experiments with lower mobility samples.
Another interesting result regarding the peak
is the distorted profile it presents that never shows up well-centered
in agreement with experiments.
In Fig. 6 we present similar results as in Fig. 5a, for a
three-component cat state. Now the resonance
peak rises at $3w_{c}=w$.
Summing up,   we have introduced the quantum
superposition of coherent states giving rise to Schrodinger cat states (even and odd)
 for ultra-high mobility samples.
Based on them, we have explained the experimentally obtained
magnetoresistance resonance
peak shift to  $2w_{c}=w$.
  In the same way, we have explained the dramatic
magnetoresistance drop that shows up in this kind of samples, irradiated or in the dark.
We have generalized the model introducing three-component Schrodinger cat states and
predicted that the resonance peak will further shift to  $3w_{c}=w$.
Finally, we have proposed, ultra-high mobility 2DES under low $B$ and $T$, as
a promising bosonic mode-based platform for quantum computing.

This work was supported by the MCYT (Spain) grant PID2023-149072NB-I00.

\end{document}